# Anonymity Network Tor and Performance Analysis of 'ARANEA' – an IOT Based Privacy-Preserving Router

*AKM Bahalul Haque, Sharaban Tahura Nisa, Md. Amdadul Bari, Ayvee Nusreen Anika*

*bahalul.haque@northsouth.edu, tahuranisa@gmail.com , amdadulbari@gmail.com, nusreen.anika@gmail.com*

## Abstract

*There was a time when the word "security" was only confined to the physical protection of things that were valuable which must be guarded against all the odds. Today, in a world where people can do things virtually have emerged the necessity to protect the virtual world. Every single facet of our life is being controlled by the internet one way or another. There is no privacy in the cyberspace as the data which we are browsing on the internet is being monitored on the other side by someone. Each work we are doing on the internet is getting tracked or the data are getting leaked without consent. To browse the internet securely we developed a router named Aranea which relates to the browser Tor. Tor gives traffic anonymity and security. The Tor browser can be used in both positive and negative purpose. Tor encrypts data, it hides the location and identity of the user, it hides the IP address of the device, it hides the network traffic and many more. By using Tor browser each user can browse the internet safely in the cyber world. Our goal is to create an additional security bridge through the router Aranea for every user so that each user can simply browse the internet anonymously.*

## Introduction

Privacy on the internet is a major concern nowadays. In traditional routing, user identity is revealed. The user activity can be tracked. An entity with enough sophisticated hardware and software can enjoy the benefits of eyeing a user on the internet always. Personal information like device information can also be tracked and intercepted. Users IP address, mac address, browser information, the browser information, and browser cache can also be collected, which is a serious violation of users privacy [1]. There are secure communication schemes available which use secure routing protocols and encryption technique for message transfer[2], but the anonymity is not ensured. Moreover, these applications are only for exchanging messages, not regular web browsing.

Considering these facts some people have rather chosen the anonymous communication as their medium as it facilitates a lot of advantages like privacy and security, not allowing the web history to be visible to others and above all track the specific behavior of the user. It is like a virtual tunnel, through which people can surf the internet without being tracked or being their surfed information revealed others. It can also be used use in instant messaging applications without disclosing the identity. In fact, it is like a guardian angel for the people who want to protect their privacy and anonymity online. Tor provides an extra layer of anonymity by routing the traffic in a random manner through routers all over the world which provides privacy. Tor encrypts data, it hides the location and identity of the user, it hides the IP address of the device, it hides the network traffic and many more.

However, to use Tor browser we need to install and configure this application in our desktop or laptop. Without installing the software, it is not possible to browse in Tor. In this cohort, it is often seen we do not use one gadget all the time. When we are in our home or outside, we use the internet on our notebook, laptop, desktop, mobile phone, tab, etc. connecting through the Wi-Fi router. It is a matter of trouble to install Tor browser in each gadget and use the internet. For that reason, to use internet smoothly along with that to protect the identity and to keep activity private of the users we created a portable router named Aranea which will provide Wifi and it will be also connected to Tor network. Aranea will help the user to browse the internet safely in anyplace, anywhere at any time as it is portable.

In this paper, the working methodology, history, and weakness of TOR will be discussed before delving deeper into designing Aranea. This will provide a comprehensive idea about the TOR network and what things should we consider while thinking about using the router.

**Historical Background**

The Tor network is followed by the onion routing. Onion routing may be called the father of Tor. The history began in the mid-90s. The journey of Tor begins in 1995 by the Military scientists at Naval research Laboratory Anacostia-Bolling military base in Southeast Washington, D.C. that can track on someone's activity on the internet anonymously. Early developers are Paul Syverson, Michael Reed, and David Goldschlag. The method was called Onion Routing which can redirect traffic into a parallel peer to peer network and bounce it randomly before sending it to the destination. It was financed by the Naval Research Office and Defense Advanced Research Projects Agency.

The primary purpose of onion routing wants privacy rather than allowing Intelligence and military personnel to scoop on someone's internet activity being unmasked.

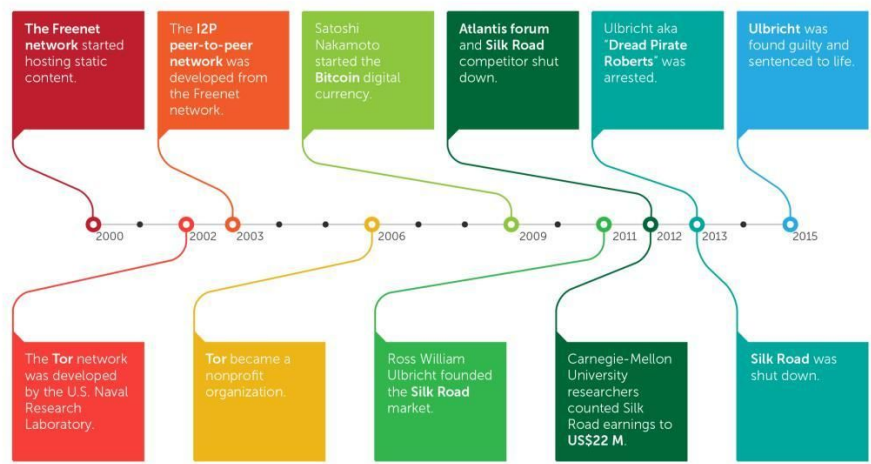

Figure 1 Timeline of deep wave implementation

The first design doesn't include any mixing but in the middle of spring 1996 Real-time mixing was added to the design. The same year on May 31 at the First Information Hiding Workshop, a proof of concept prototype on Solaris 2.5.1/2.6 was deployed. The prototype consists of a 5 node system running on a single machine at NRL with proxies for Web browsing which features with and without importance of the application protocol data. It also includes tryst points and tagging attacks.

Design utilizing Onion Routing for location anonymous use of mobile phones and other location tracking devices is published at the Security Protocols Workshop in April 1997.

In 1998 several other networks are installed. A distributed network of 13 nodes at NRL,NRAD, and UMD was also set up. NRAD directory was also build up which runs on Windows NT and redirects all TCP traffic to the Onion Routing network without the need for special proxies. Some other proxies were also built including those for HTTP (anonymizing and no anonymizing versions), FTP, SMTP, and rlogin.

In July 2000 JAP (Java Anon Proxy) Web Mixes goes online in autumn. TU Dresden organized it as a mixed cascade based Web proxy. The threat model and approach are more based on traffic from a fixed group or mixed independent operators, while Onion Routing includes elements of the path and jurisdictional uncertainty on a per circuit basis. Tor network deployment of and releasing of Tor code under the open and free MIT license is done around October 2003. By the end of 2003, Tor network has about a dozen volunteer nodes one in Germany and the rest in the US. The JAP team independently implemented a client for Tor in 2004 that functions with the Tor network.

Location hidden services are employed in the spring when the hidden wiki is set up. Funding from ONR and DARPA ends in 2004. Funding from EFF for continued Tor deployment and development begins. Internal NRL improvement for location hidden servers was still on the run. By the end of 2004, there are over 100 Tor nodes on three continents [3][4][5].

**Goals of Using Tor-based Router**

The goal of almost every anonymous network is to surf the World Wide Web anonymously. Anonymity over the internet provides safe browsing. Here safe browsing alternates by the term protected and private. Moreover, tor only provides security and privacy but it also provides the facility to host web services with hidden services. Silk Road is a good example of this. These kinds of websites are operated outside the extensive surveillance of the state government of almost all over the world. As described before Tor network is the ultimate supplementary tool for encryption and security.

For those people who want to express their valuable opinion over the internet where the concerned persons country or state does not allow to speak. Journalist and bloggers are generally seen to be of these kinds. On the other hand activists and terrorists use this network to go on with their heinous crimes. Brief description of what kind of people use Tor and for what purpose are also discussed in the following section.

**Normal People**

They protect their identity and individuality from some crooked persons marketers and identity thieves. Sometimes it is seen that the ISPs use users browsing history to earn money. Though they promise anonymity of their browsing history to the users it is not the case always. They sell the browsing logs at very little to a high amount. They safeguard personal data from irresponsible corporations. This is done where privacy is seldom breaches and/or betrayals of private data. Right from the lost back up to the research data given for research nothing is safe in this case and is on the verge of great loss.

Parents use Tor to safeguard their children. Sometimes children use the internet for their study purpose or entertainment but malicious people use this facility to track the location commit crimes. In this case tor can be used as a powerful weapon against them.

Research sensitive topics are also transferred over the internet using this technology so that it is not leaked to the wrong people. Often safe browsing also raises the flags for unlawful activities. In that case, the Tor network can be used to conceal the physical identity of the user. Moreover in some countries, there are some blocked websites like Facebook or YouTube. In this case, Tor can be used to access these sites

**Journalists and Law Enforcement Agencies**
Journalists without Borders use to travel to different countries and track internet prisoners. In this case, they are advised to use Tor for anonymity with respect to the location and the identity. US international Broadcasting Bureau supports this anonymous network for private and safe access to media.
To facilitate officials to use different questionable websites anonymously and without being tracked, Tor is a great friend to them. Investigations can be hampered if the system usage log is found in these type of websites. To operate undercover/sting operations (online) law enforcement officials use Tor.

**Activists and Whistleblower**
Activists use Tor to anonymously send reports about different contradictory and unlawful activities from the danger zone. In this case, Human Rights activists are a good example. Some environmental groups use Tor to continue their activities while they are kept under surveillance. This helps to hide their identity. Global Voices (Citizen Media stories around the world), recommends using Tor to the bloggers.

**High and Low Profile People**
Being a public spotlight it sometimes become a curse. High profile people faces problem in their online private life. Their opinions have a great impact to the society. In this Tor provides the facility to conduct their activities online. Anonymity gives voice to voiceless. Sometimes people living under the poverty line or working in a lower position in a company wants to express their views and ideas but hesitates for the fear of losing his job. Tor, in this case, extends a helping hand.

**Business Executives**
To study the online market for starting a new business and also analyzing the marketing strategy and other information about the competitors over the internet is a very common method in business. In this case, some business websites are set to give misleading information to the users dependent on the location. Tor provides the facility to use the internet independent of location tracking. To keep economic and financial strategies confidential it is important to hide the research activities that are done by the analysts. In this case, to hide the log Tor network is important too.

**Working Principle**
Tor network distributes the internet data package through its sequence of nodes form the source of the packet to the destination. Those tor nodes are independent and cannot know the whole path of the data packets. So no one can point the link of the source and destination. At the same time to confirm the privacy and security tor user selects random nodes from the tor network system to firm the pathway from source to destination. After some time Tor users change selected nodes to set up a new pathway to keep the pathway form monitoring. Three kinds of tor nodes exist in the anonymous communication of the Tor network. Server node can only visit the destination websites. It is also capable of relaying other data packets of the tor node.

The server is able to initiate tor router and as soon as it is done it can be the mediate relay node for all other tor routers in the network. The other kind of tor network is a client node. It can only begin the tor router, but not able to relay the nodes. Another type of tor node is known as the directory server. It stores the information of the server nodes available in the tor network.

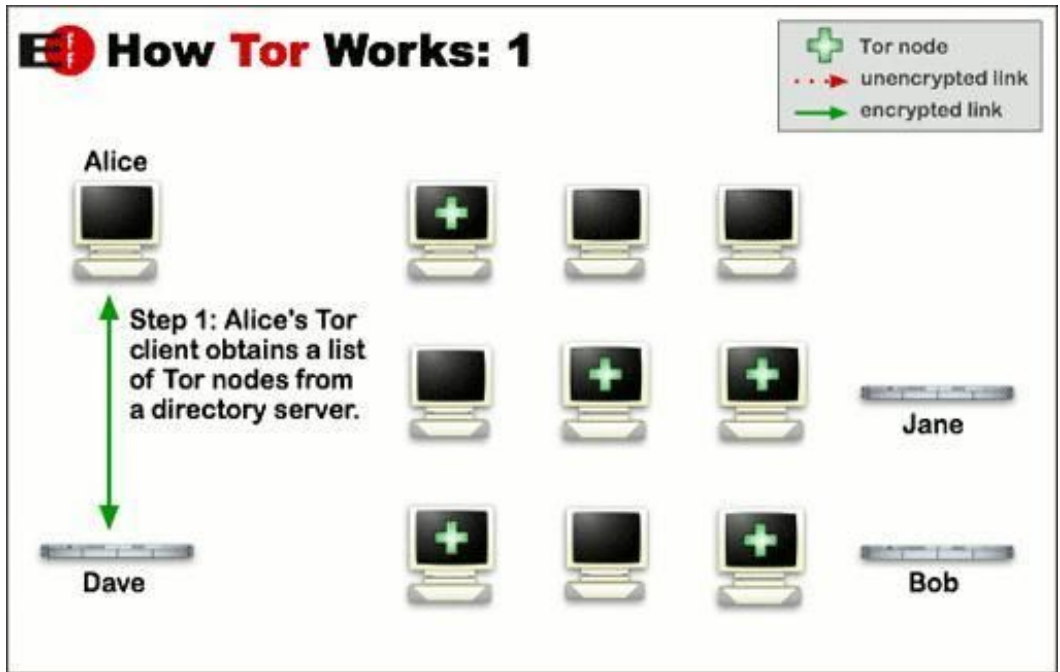
Figure 2 Step 1 of Tor Routing

As elastration of the above figure, Alice is a tor client and wants to visit destination Bob which is a website. Between Alice and Bob, there are many Tor server nodes like Dave. For Alice to communicate to the website Bob, Alice needs to collect the required information about Tor's server nodes. Here the directory server Dave meets this requirement. As soon as Alice finds the required data about Tor's server it can start to establish the circuits towards the destination.

Now the client node has got the necessary information about server nodes. It's now time to build circuits. The client node establishes a Tor circuit incrementally with 3 other server nodes. The circuit set up is completed in a step by step process with the help of TLS and private key. The extension of the circuit is done as one hop at a single time. Every server node and the circuits know only the information of the previous node. All other information is not delivered to the following nodes. After finishing the circuit set up the client node begins to transmit the data packet. Several data streams can be transmitted through the circuits as HTTP stream request. So without disclosing the sensitive information like IP address or data packet size to the authority, the client node can successfully transmit the required data in an absolutely safe and secure process.

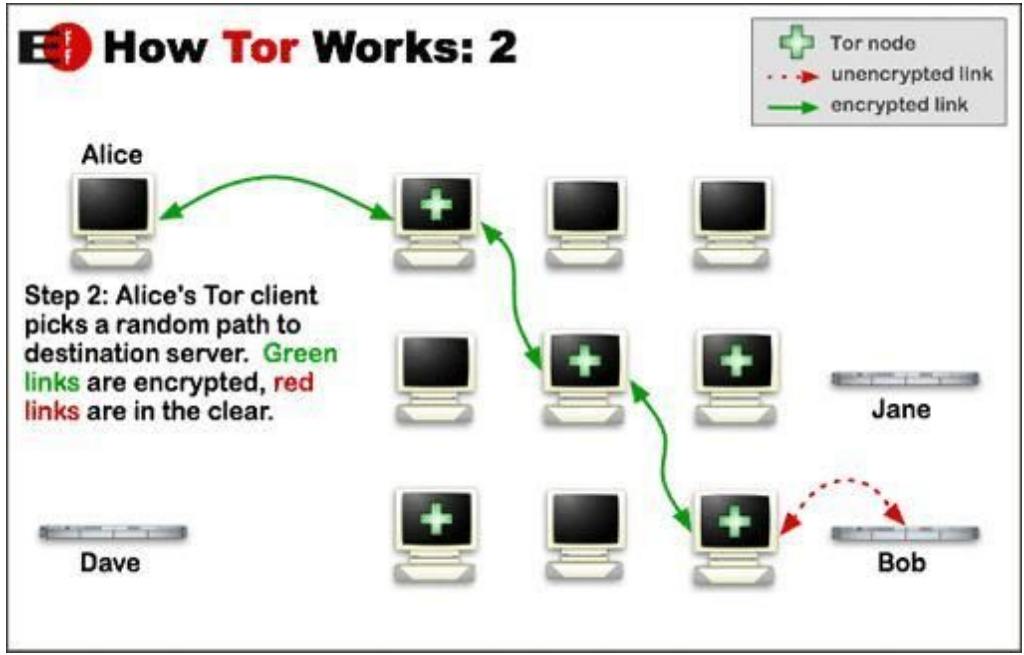

Figure 3 Step 2 of Tor Routing

The data packets transmitted by the client node are like a three-layer onion router. A single hop through the circuitry only can wrap a single layer with its private key. Every hop through the Tor circuitry only knows its predecessor and successor node.

Data packet structure of the Tor network described in the following section is illustrated with the help of the onion layer. In the Tor circuit, the final mediate node (server node) is defined as exit node. As described earlier through the entire path the data packed is secured through a transport layer security protocol. This is said as one of the weaknesses of the Tor communication which helps sniffers to collect sensitive information by the detailed address cannot be obtained in this method. Please refer to the picture for better illustration. To protect anonymity, in other words, to be highly efficient in its working procedure

To build the circuit, through which the data packet is transmitted is changed in every ten minutes.

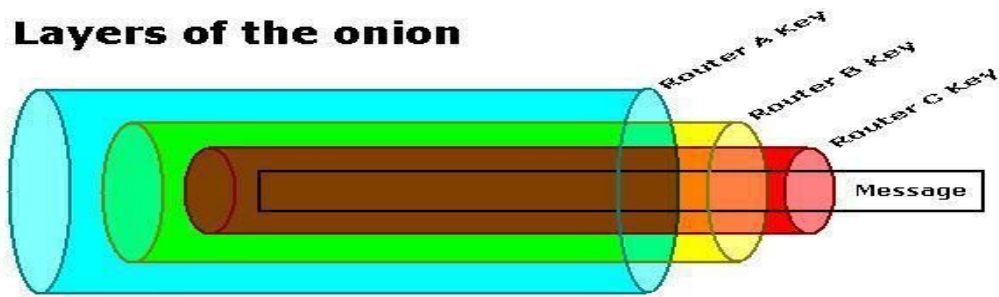

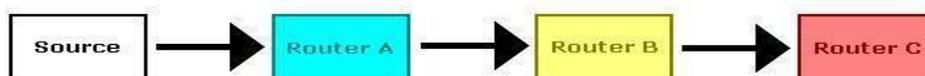

Figure 4 Data packet structure

Moreover for safeguard from the sniffers and/or to prevent from monitoring the network it also does the same in every ten minutes. As

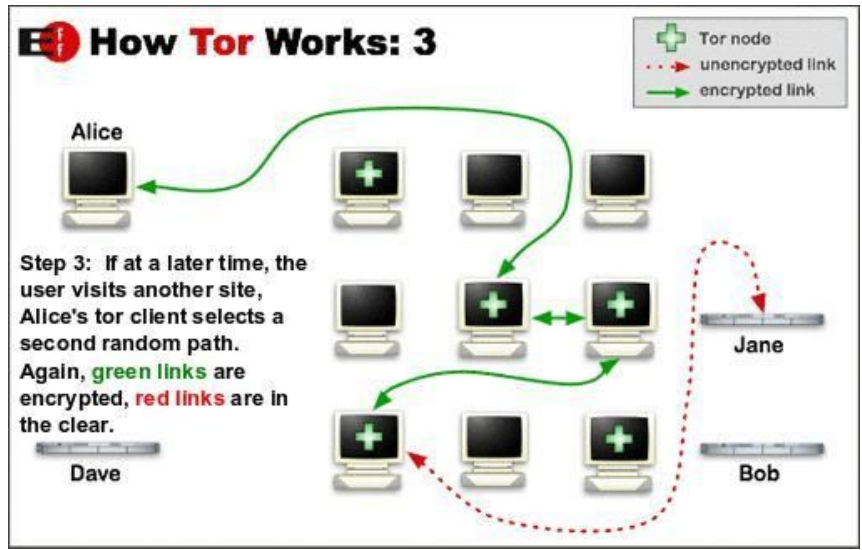
Figure 5 Step 3 of Tor Routing

in the figure, while Alice sends data packet for the first time it uses a circuit consists of three nodes. If Alice wants to visit another website now, Tor will set up a new circuit path for her to protect its anonymity. For example now if Alice wants to visit Jane the circuit path is changed (see Fig 4). Now Alice will transmit data through this path. For anonymity, over the internet, the Tor network uses this methodology in general. It achieves anonymity and protects through such types of twisted and distributed route. In fact, Tor can be defined as an effective tool for payload encryption while transmitting data.

Anonymous communication systems are designed to protect users from malicious websites and network eavesdroppers by providing means to hide the content and metadata of communication. Although Tor hides the routing information and communication content, the analysis of the network tariff alone may be very informative to the attacker with sufficient capabilities. The attacker tricks a user by sending a distinct file by the web browser which can be detected by traffic analysis. Both the attack and traffic analysis can be performed by an adversary with limited resources. Traffic analysis or Flooding method and Attack are mentioned here with brief details [7][8].

## Related Work
Roger Dingledine, Nick Mathewson, Paul Syverson presented Tor, a circuit-based low –latency anonymous communication service [9]. Tor works on the real-world internet, requires no special privileges or kernel modifications, requires little synchronization or coordination between nodes and provides a reasonable trade-off between anonymity, usability, and efficiency. Onion routing is a distributed overlay network designed to anonymize TCP-based application like web browsing, secure shell, and instant messaging.

There have been numerous works related to TOR or in the field of computer networks security after TOR. Expanded packet processing system which appends monitoring subsystems to verify accurate operation. Security Packet Processing Platform (SPPP) architecture detect an attack via monitoring tool and a retrieval subsystem to limit footprint. "Design a secure Router System for Next-generation Networks" [10] which presented the design of a Secure Packet Processing Platform (SPPP) that can protect computer router systems. They used an instruction-level monitoring system to detect deviations in processing behavior.

Other works or surveys related to TOR are how to improve that software system and avoid traffic and identified key weaknesses in the design of Tor, the most used anonymous communication network. As the number of Tor user increases, the performance of Tor degrades badly due to traffic is not fairly distributed. As a result, the Tor network does not provide a better quality of communications between users. "Improving the TOR Traffic Distribution with Circuit Switching Method" proposed a simple method of dynamic circuit switching on the Tor application level to distribute the bulk and light traffic on Tor [11].

Michael Alsabah's "Performance and Security Improvements for Tor: A Survey" focused on the points which should be given proper importance in order to improve the performance and services of TOR [12]. Some notable works in communication field are "Communication network with secure access for portable users" by Naresh Chand and Bruce M. Eteson, which invented a network that allows a portable user to engage in wireless communications wherein normal messaging is routed over an RF link with the user and classified or other highly sensitive messages are contained over a secure FSO link that can be established by or with the user when desired [13]. Another work is "Wireless access point apparatus and method of establishing secure wireless links" by Katsuhiko Yamada and Azuma Tsubota. One of their invention's object is to provide wireless access point apparatus that can establish secure wireless links between access points in a wireless mesh network by using a separate encryption key for each link [14].

**Tools and Technologies for Design and Development of Aranea**
- TOR
- Dnsmasq
- Hotspot
- Raspberry Pi
- Modem

TOR: Tor is free and open-source software for enabling anonymous communication [15]. It gives user anonymity and data security [16]. Dnsmasq: It is a tool which provides Domain Name System (DNS) forwarder, Dynamic Host Configuration Protocol (DHCP) server, router advertisement and network boot features for small computer networks [17].

Hotspot: Hotspot is a small daemon to create a Wi-Fi hotspot on Linux. It depends on hotspot for AP provisioning and DNS mask to assign IP addresses to devices [18].

## 4.System Architecture

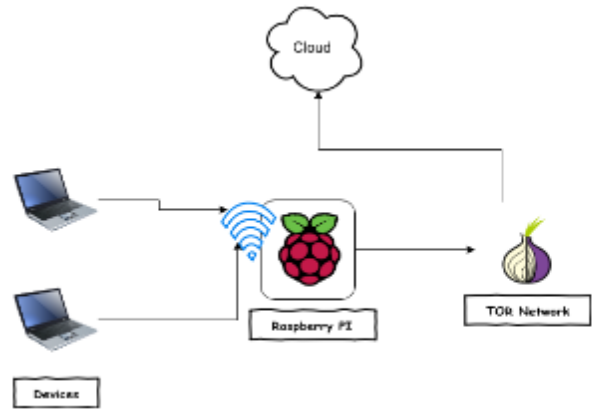

Figure 6. System Architecture of Aranea

The router is based on a raspberry pi 3. It has one GSM modem with it for the internet connection and a 2000MaH power bank for power supply and a strong waterproof case. The core OS of the device is Raspbian. This OS is used to make the device lightweight. The raspberry pi was configured to act like a normal router. The device broadcast SSID using the Wi-Fi module. It has also run a DHCP server in it, which accepts and assigns IP to the devices that connect to it. It also gives DNS service using dnsmasq. Basically, it acts as an access point with WPA2/PSK encryption. It has tor software installed and configured. When the device boots up the TOR software and all other things for the access point run automatically. When any traffic goes out using the device it uses the TOR network because an interface of the device is connected to the TOR network. Tor gives traffic anonymity and security.

**Results**

Internet Speed

Internet speed is observed using the normal router and using the "Aranea" router. It is found that Aranea router slows down the internet speed.

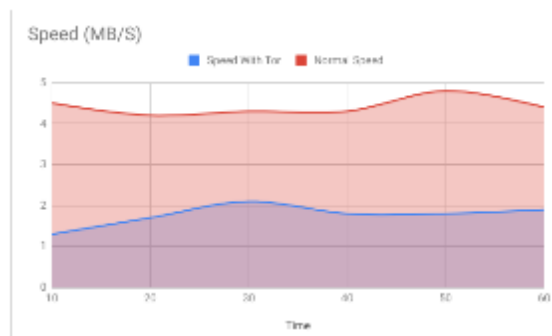

Figure 7. Internet speed of Aranea Router

**Latency:**
It also increases latency. We observed that latency increases 5 times more when we use "Aranea" router.

**DNS Leak:**
DNS leak is also tested [19] on the "Aranea Router". It can hide DNS, so DNS request will also be not revealed.

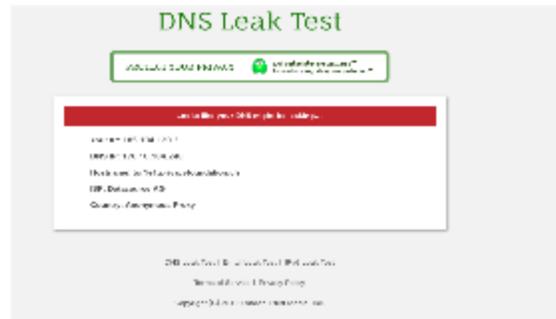

Figure 8. DNS Leak test of Aranea Router

## 6.Benefits

To browse the internet safely and securely we developed a portable router named Aranea which will relate to the tor browser. There are many benefits to using Aranea. Among them, the five main benefits are anonymity, data security, portable, cost efficiency, simplicity to use.

**Anonymity:**
Internet security is not less important than we often share our personal views, personal pictures in social media. With or without our information our personal data are getting leaked. People are using our data's for making money some are doing business on the other side some are trapping innocence due to lack of anonymity. Internet is a platform for the users where they can share their thought, raise their voice, speak without any fear but due to lack anonymity, people don't get the freedom to use this platform safely in the cyber world. Online anonymity plays a very important role in freedom of expression. In our daily life due to lack of online anonymity, thousands of people per day are facing the fear of identity protection, personal harassment and many more. The router Aranea creates an anonymity network with the help of Tor which enables other users to access the web while blocking any tracing the users' identity on the internet. Using Aranea makes it more difficult to trace Internet activity to the user. The anonymity networks of Aranea prevents traffic analysis and network surveillance.

**Data security:**
It is not possible in social media to secure your data. Whatever Data we are giving in the social media each data is getting leaked under one's nose without any sort of clue. Securing data is very important for any user in the world of cyberspace. Aranea will help a user to prevent unauthorized access to computers, databases, and website. Aranea, encrypt data of the user. It will be tough for a hacker to collect data whoever is using the router Aranea as it is creating network traffic to hide data.

**Portable:**
The router Aranea has very, therefore, it is easy to carry. We can take the router Aranea anywhere any place we want as it is portable. Users can use the internet without any terror in the cyber world by using router Aranea.
Cost Efficient: To create the router Aranea we used raspberry pi 3. The core OS of the device is Raspbian. This OS is used to make the device lightweight. The raspberry pi was configured as a normal router. The cost is cheaper

**Simplicity:**
personal security as our personal life is browsing the internet safely. deeply linked with it. While browsing the internet we
The router Aranea is very simple and easy to use. Tor browser is connected through the router Aranea. The device is needed to connect with Ethernet or Internet Modem to create a network. It is securing the Wi-Fi network for the users. The router provides all the services that a normal router gives. The router Aranea is portable we can use internet easily and reliably anyplace anywhere we want by connecting devices such as computers, laptop, notebook smartphone, etc. The router Aranea is creating an additional security bridge for every user so that each browser can simply

## Limitations
**Slow internet speed:**
Tor network is slow compared to the normal network. Tor gives traffic anonymity. For traffic anonymity, it bounces via other computers in the various part of the world. Tor provides anonymity by building circuits relays. Instead of connecting directly to the destination server, it creates other connection between the relay of the circuit, and this takes more time. The Tor relays are run by volunteers in a decentralized way due to that tor has slow internet speed.

**Illegal to use in some countries**
Browsing Tor is not illegal as it is free and open source software. It is used for enabling anonymous communication. Browsing the internet via Tor in dark websites [20] is illegal, doing illegal activities such as dealing drugs, making pornography putting them on dark websites can put you trouble on the cyber world. There are several countries that block VPN Saudi Arabia, Bahrain, Iran, and China. China is the only country to successfully block Tor [21].

**Possible Attack on Aranea**

**Tor flooding**
Murdoch-Danezis clogging attack based on flooding attack and looking for delays in the connection and using network delays as a potential method of identifying the sender. In this approach an attack that allows a single malicious tor server and a colluding web server or in other words client server to identify all three tor nodes of a Tor circuit used by a client by a given session. Generally, the exit nodes identity can be known to the service provider. But this system cannot identify the client directly only the entry node into the tor network can be known. The approach is as follows when a client tries to connect with the malicious web server. The server then controls its data transfer to the client in such a way that to make the traffic pattern easily identified by the observer. In this method at least one Tor server is controlled by the timing circuits among other Tor servers. By sending traffic through that timing circuit it is possible to detect which tor servers are showing a pattern like that the attacker web server is generating. As Tor does not reserve bandwidth for each connection, one connection experience in a tor node is heavily loaded and other experiences increase latency. By locating which node in the tor network exhibits the attacker generated signals the entire tor network used by the client can be mapped.

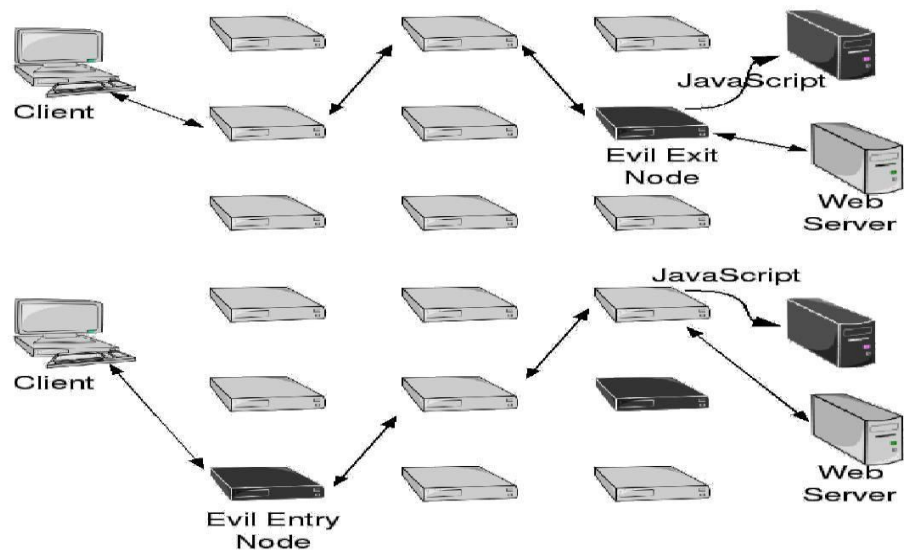

Figure 9 Tor flooding technique

Just like the figure above a malicious exit node modifies web pages, inserting JavaScript code that frequently connects to a logger server, sending a idiosyncratic signal along the link (top). If the client then uses a malicious entry node while that JavaScript is still on the run, the entry node can detect the signal, and the attacker can thus accompany the client with his communications on the same machine.

**Fingerprint method**
The main objective of an observer in a typical Website Fingerprinting scenario is to identify which page the user is visiting. The observer may want to learn this information for surveillance or intelligence purposes.

The Website Fingerprint attack is typically treated as a classification problem, where classification categories are web pages and observations are traffic traces. The observer first collects traffic traces by visiting web pages and trains a supervised classifier using length, direction and inter-arrival times of network packets. When a user visits a webpage with Tor, the observer can track the users by intercepting the LAN or by controlling the entry point of the tor networks or by accessing the routers of users ISP. He then runs the classifier on the intercepted network trace to guess the site the user has visited.

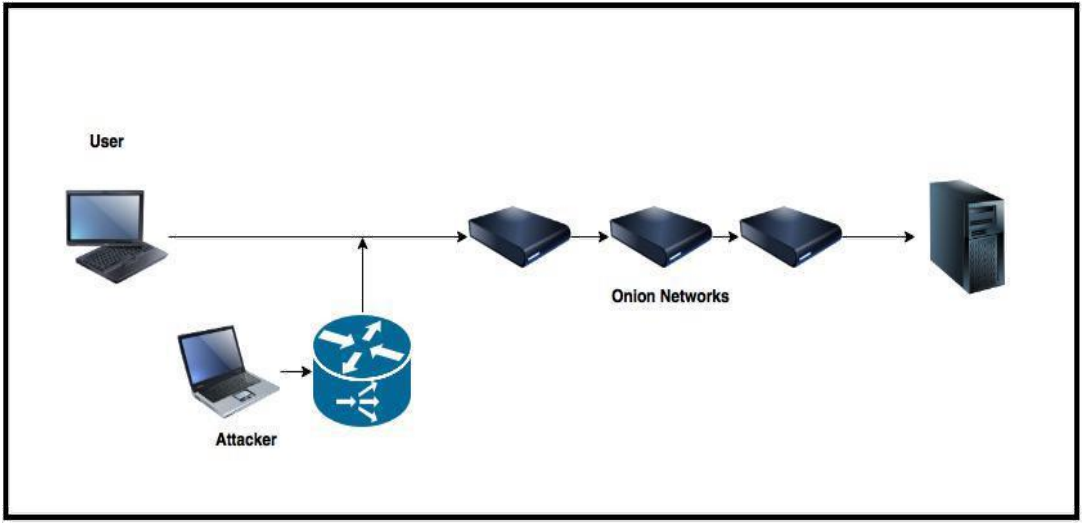

Figure 10. the basic Website Fingerprint targeted attack in Tor.

As shown in Figure 1 attacker taps between the user and the tor entry gate and collect information. Then the observer compares it with the blueprint of the web page and distinguishes two types of attack one is Targeted and the one is Non-targeted.

In the first kind observer targets a specific victim to retrieve his browsing activity. This allows the attacker to train a classifier under conditions similar to those of the victim (see Figure 1), potentially increasing the success of the attack. The observer may have enough background knowledge about the user to reproduce his configuration, or he could detect it from the observed traffic data.

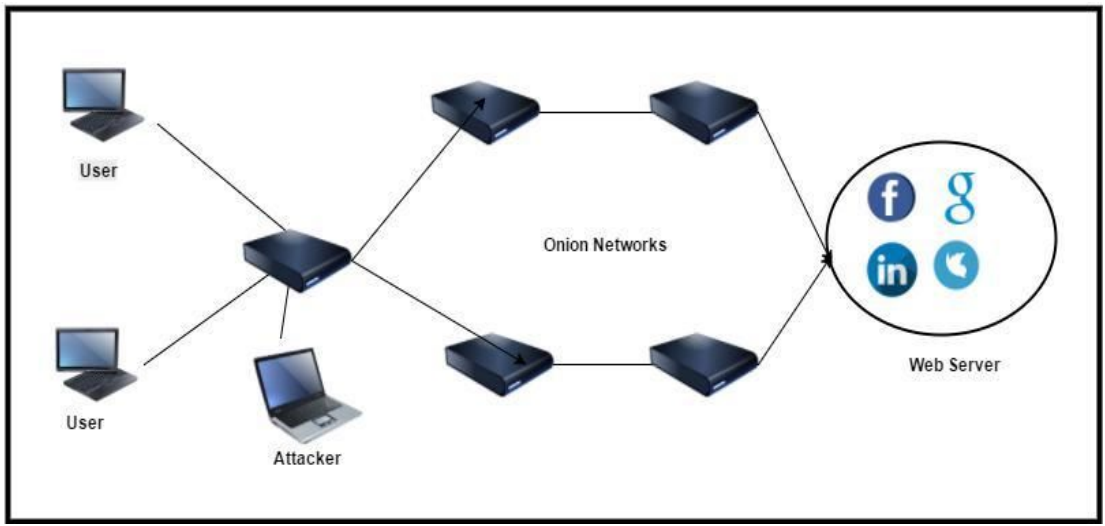

Figure 11. Malicious entry guard.

In Non-Targeted surveillance, the observer targets a set of users instead of one. ISPs, routers and entry point operators are in a position to make this attack as they can intercept the network traffic of many users (see Figures 10,11 and 12 respectively).
The attacker trains the classifier on a specific setting and uses the same classifier on all communications that he observes.

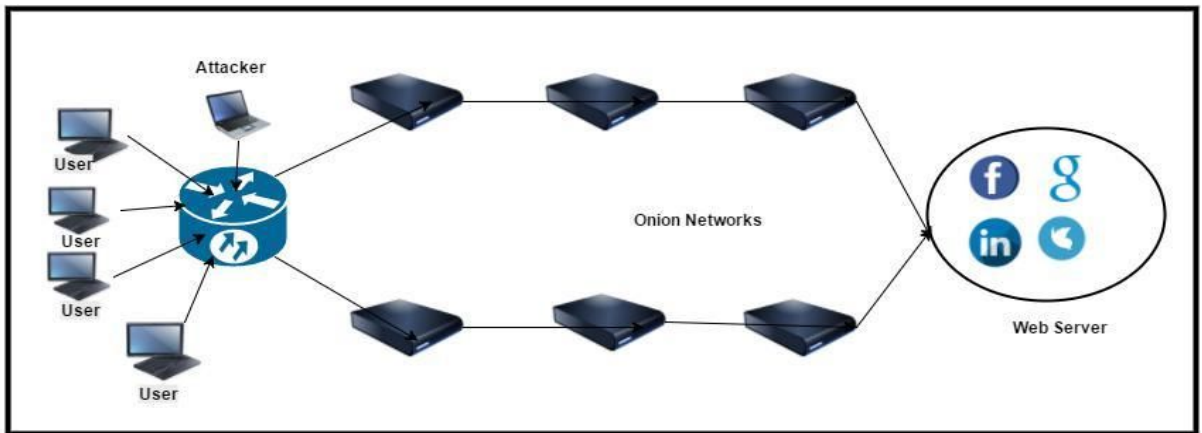

Figure 12. ISP level adversary

**Conclusion**

In the Enlarging globe of the internet in each footfall of our life, we need to be vigilant about our activities in the cyber world. People are becoming conscious these days to protect themselves from internet threats. The primary objective of this work was to inspect the privacy solution and create a platform for common users to keep their data private with cost effacing. The router Aranea was successfully able to veil user data from third parties and its camouflaged ISP. The router is portable, and it can be easily connected via Ethernet or WIFI. The router is cost-efficient and effective. However Fundamental target of the project is achieved by creating a secure platform in the virtual world for the users.